\documentclass[10pt, conference, compsocconf]{IEEEtran}
\ifCLASSINFOpdf
\else
\fi
\usepackage{subfigure}
\usepackage{graphicx}
\usepackage{amsmath}
\usepackage{wrapfig}

\begin{document}
%
% paper title
% can use linebreaks \\ within to get better formatting as desired
\title{Multislice Modularity Optimization in Community Detection and Image Segmentation}

% author names and affiliations
% use a multiple column layout for up to two different
% affiliations
\author{\IEEEauthorblockN{Huiyi Hu, Yves van Gennip, Blake Hunter, Andrea L. Bertozzi}
\IEEEauthorblockA{Department of Mathematics\\
University of California, Los Angeles\\
Los Angeles, CA, USA\\
Email: huiyihu@math.ucla.edu, yvgennip@math.ucla.edu,\\ blakehunter@math.ucla.edu, bertozzi@math.ucla.edu}
\and
\IEEEauthorblockN{Mason A. Porter}
\IEEEauthorblockA{Oxford Centre for Industrial and Applied Mathematics,\\ Mathematical Institute;\\ and CABDyN Complexity Centre
\\
University of Oxford, Oxford, UK\\
Email: porterm@maths.ox.ac.uk}
}

% conference papers do not typically use \thanks and this command
% is locked out in conference mode. If really needed, such as for
% the acknowledgment of grants, issue a \IEEEoverridecommandlockouts
% after \documentclass

% for over three affiliations, or if they all won't fit within the width
% of the page, use this alternative format:
% 
%\author{\IEEEauthorblockN{Michael Shell\IEEEauthorrefmark{1},
%Homer Simpson\IEEEauthorrefmark{2},
%James Kirk\IEEEauthorrefmark{3}, 
%Montgomery Scott\IEEEauthorrefmark{3} and
%Eldon Tyrell\IEEEauthorrefmark{4}}
%\IEEEauthorblockA{\IEEEauthorrefmark{1}School of Electrical and Computer Engineering\\
%Georgia Institute of Technology,
%Atlanta, Georgia 30332--0250\\ Email: see http://www.michaelshell.org/contact.html}
%\IEEEauthorblockA{\IEEEauthorrefmark{2}Twentieth Century Fox, Springfield, USA\\
%Email: homer@thesimpsons.com}
%\IEEEauthorblockA{\IEEEauthorrefmark{3}Starfleet Academy, San Francisco, California 96678-2391\\
%Telephone: (800) 555--1212, Fax: (888) 555--1212}
%\IEEEauthorblockA{\IEEEauthorrefmark{4}Tyrell Inc., 123 Replicant Street, Los Angeles, California 90210--4321}}

% use for special paper notices
%\IEEEspecialpapernotice{(Invited Paper)}

% make the title area
\maketitle

%======================================================================================================================
%Abstract
\begin{abstract}

Because networks can be used to represent many complex systems, they have attracted considerable attention in physics, computer science, sociology, and many other disciplines.  One of the most important areas of network science is the algorithmic detection of cohesive groups (i.e., ``communities") of nodes.  In this paper, we algorithmically detect communities in social networks and image data by optimizing multislice modularity. A key advantage of modularity optimization is that it does not require prior knowledge of the number or sizes of communities, and it is capable of finding network partitions that are composed of communities of different sizes.  By optimizing multislice modularity and subsequently calculating diagnostics on the resulting network partitions, it is thereby possible to obtain information about network structure across multiple system scales. We illustrate this method on data from both social networks and images, and we find that optimization of multislice modularity performs well on these two tasks without the need for extensive problem-specific adaptation.  However, improving the computational speed of this method remains a challenging open problem.

\end{abstract}


% For peer review papers, you can put extra information on the cover
% page as needed:
% \ifCLASSOPTIONpeerreview
% \begin{center} \bfseries EDICS Category: 3-BBND \end{center}
% \fi
%
% For peerreview papers, this IEEEtran command inserts a page break and
% creates the second title. It will be ignored for other modes.
\IEEEpeerreviewmaketitle

%======================================================================================================================
% Methods
\section{Methods}\label{methods}
% no \IEEEPARstart

Many networks can be partitioned into \emph{communities}, such that they consist of cohesive (and often dense) groups of vertices with sparse connections between distinct groups \cite{comnotices}.  Perhaps the most popular way of detecting communities algorithmically is by optimizing the quality function known as modularity \cite{newman}:
\begin{align}\label{mod}
	Q= \frac{1}{2m}\sum_{ij} \left(A_{ij}-\gamma \frac{k_i k_j}{2m}\right)\delta(g_i,g_j)\,,
\end{align}	
which measures how well a network can be partitioned into disjoint groups of nodes.  In (\ref{mod}), $A_{ij}$ are the elements of the graph's adjacency matrix ${\bf A}$, the sum of all of the edge weights in the network is $m$, $k_i$ is the strength (i.e., weighted degree) of node $i$, and  the resolution parameter $\gamma$ \cite{resolution} enables us to uncover community structure at different scales.  The modularity of a network partition measures the fraction of total edge weight within communities minus that expected if edges were placed randomly according to the null model $P_{ij} = k_ik_j/(2m)$, which preserves a network's expected strength distribution.  Finding a network partition that attempts to maximize $Q$ allows one to probe a network's community structure.  In contrast to traditional forms of spectral clustering, modularity optimization requires no knowledge of the number or sizes of communities, and it also allows one to segment a network into communities of disparate sizes (even for a fixed value of $\gamma$) \cite{comnotices,newman}.  \\
%==============================================================================
\begin{figure}
\centering
%\subfigure[]{\includegraphics[width=0.235\textwidth]{schematic_n.png}}
\subfigure[]{\includegraphics[width=0.235\textwidth]{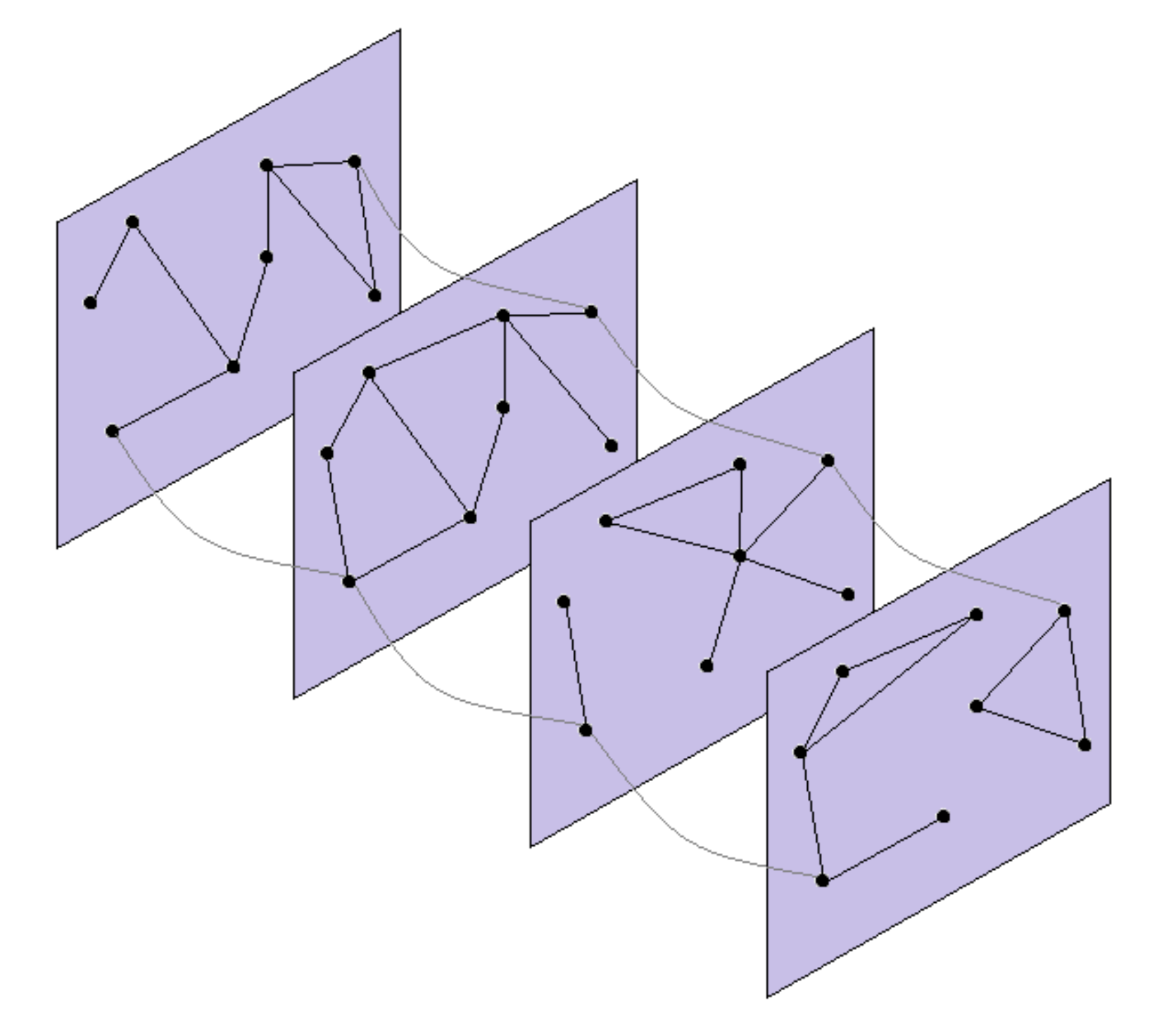}}
%\subfigure[]{\includegraphics[width=0.235\textwidth]{sample_multislice}}
\subfigure[]{\includegraphics[width=0.235\textwidth]{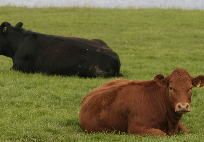}}
\caption{(a) Schematic of a multislice network. (We reproduce this image from \cite{mason} with permission from the authors.) (b) Image of a pair of cows, which we downloaded from the Microsoft Research Cambridge Object Recognition Image Database \cite{cowimage} (copyright~\copyright~2005 Microsoft Corporation). It is modified to produce the segmentation in Fig.~\ref{cow}.
}
\label{schematic}
\end{figure}
%======================================

Optimization of modularity was recently generalized to ``multislice" networks \cite{mason}, which are represented using adjacency tensors and consist of layers of ordinary networks.  The framework of multislice networks can thereby be used to represent time-dependent or multiplex networks. In Fig.~\ref{schematic}(a), we show a schematic of a multislice network.  Using this framework, we define a generalized modularity function \cite{mason}
\begin{equation}\label{multimod}
		\small{Q_{\mathrm{multi}}= \frac{1}{\mu}\sum_{ijsr} \left[(A_{ijs} -\gamma_s \frac{k_{is} k_{js}}{2m_s})\delta_{sr} + \delta_{ij}C_{jsr}\right]\delta(g_{is},g_{jr})\,, }
\end{equation}
where $g_{jr}$ indicates that community assignment of node $j$ from slice $r$, the intraslice edge strength of node $j$ in slice $s$ is $k_{js}=\sum_i A_{ijs}$, the corresponding interslice edge strength is $ c_{js}=\sum_r C_{jsr}$, and $2\mu=\sum_{jr} k_{jr}+c_{jr}$.  In (\ref{multimod}), one can use a different resolution parameter $\gamma_s$ in each slice. For a given slice $s$, the quantity $A_{ijs}$ gives the edge weight between nodes $i$ and $j$.  For a given node $j$, the quantity $C_{jsr}$ gives the interslice coupling between the $r$th and $s$th slices.\\

Optimization of the ordinary modularity function (\ref{mod}) has been used to study community structure in myriad networks \cite{comnotices}, and it has also been used in the analysis of hyperspectral
images \cite{hyperspectral} recently.
%Apart from community detection problems, the basic modularity method has been applied to hyperspectral images . 
In our work, we optimize multislice modularity (\ref{multimod}) to examine community structure in social networks and segmentation of images.  In each case, we start with a static graph, and each layer of the multislice network uses the same adjacency matrix but associates it with a different resolution-parameter value $\gamma_s$.  We include interslice edges between each node $j$ in adjacent slices only, so $C_{jsr} = 0$ unless $|r - s| = 1$.  We set all nonzero interslice edges to a constant value $\omega$.  This setup, which was illustrated using the infamous Zachary Karate Club network in \cite{mason}, allows one to detect communities using a range of resolution parameter values while enforcing some consistency in clustering identical nodes similarly across slices.  The strength of this enforcement becomes larger as one increases $\omega$. To optimize multislice modularity (\ref{multimod}), we use a Louvain-like locally-greedy algorithm  \cite{louvain,genlouvain}. 

%======================================================================================================================
% Figures

% You must have at least 2 lines in the paragraph with the drop letter
% (should never be an issue)
\begin{figure}
\centering
\subfigure[]{\includegraphics[width=0.23\textwidth]{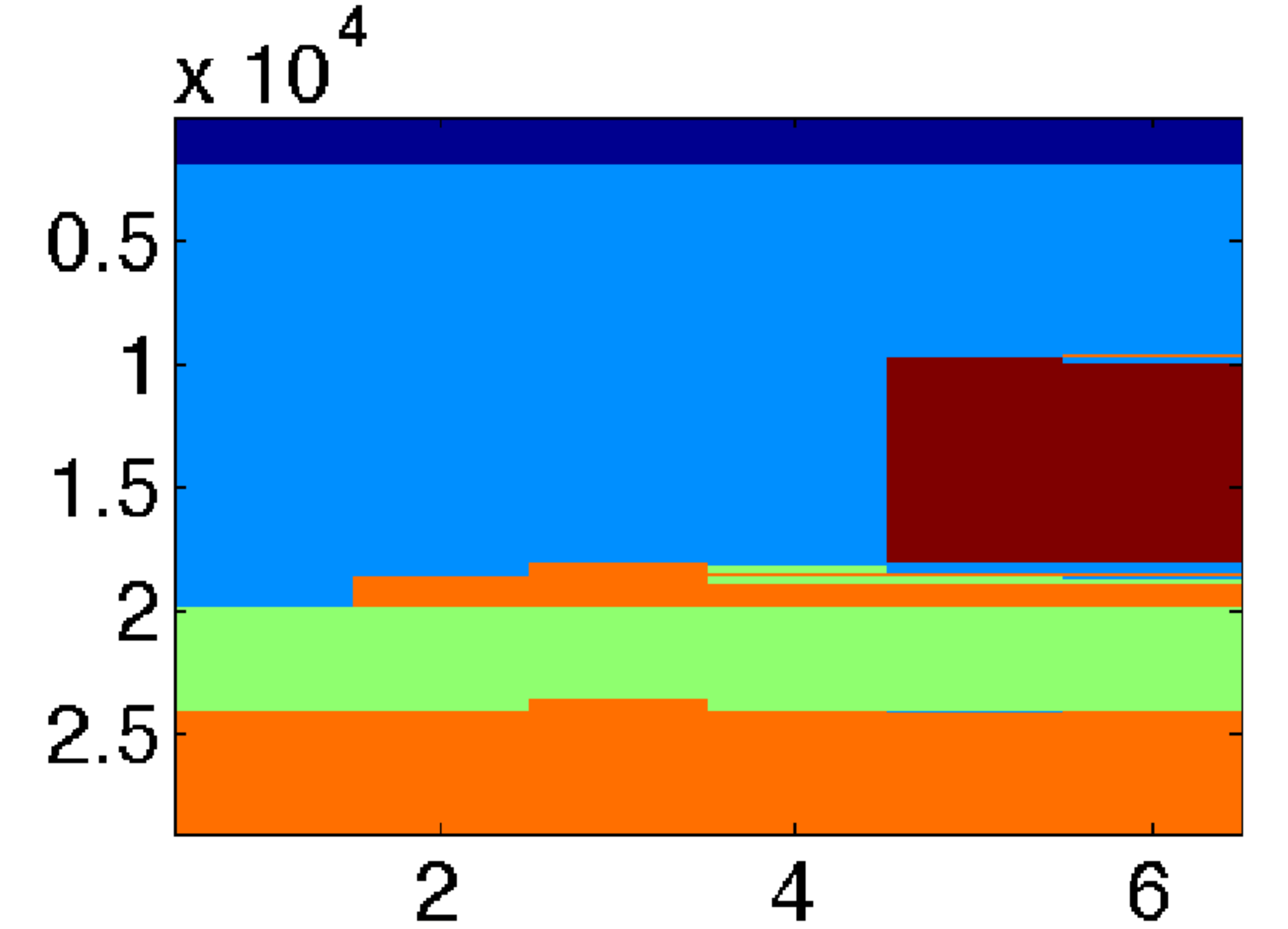}} \hspace{-3pt}
\subfigure[]{\includegraphics[width=0.23\textwidth]{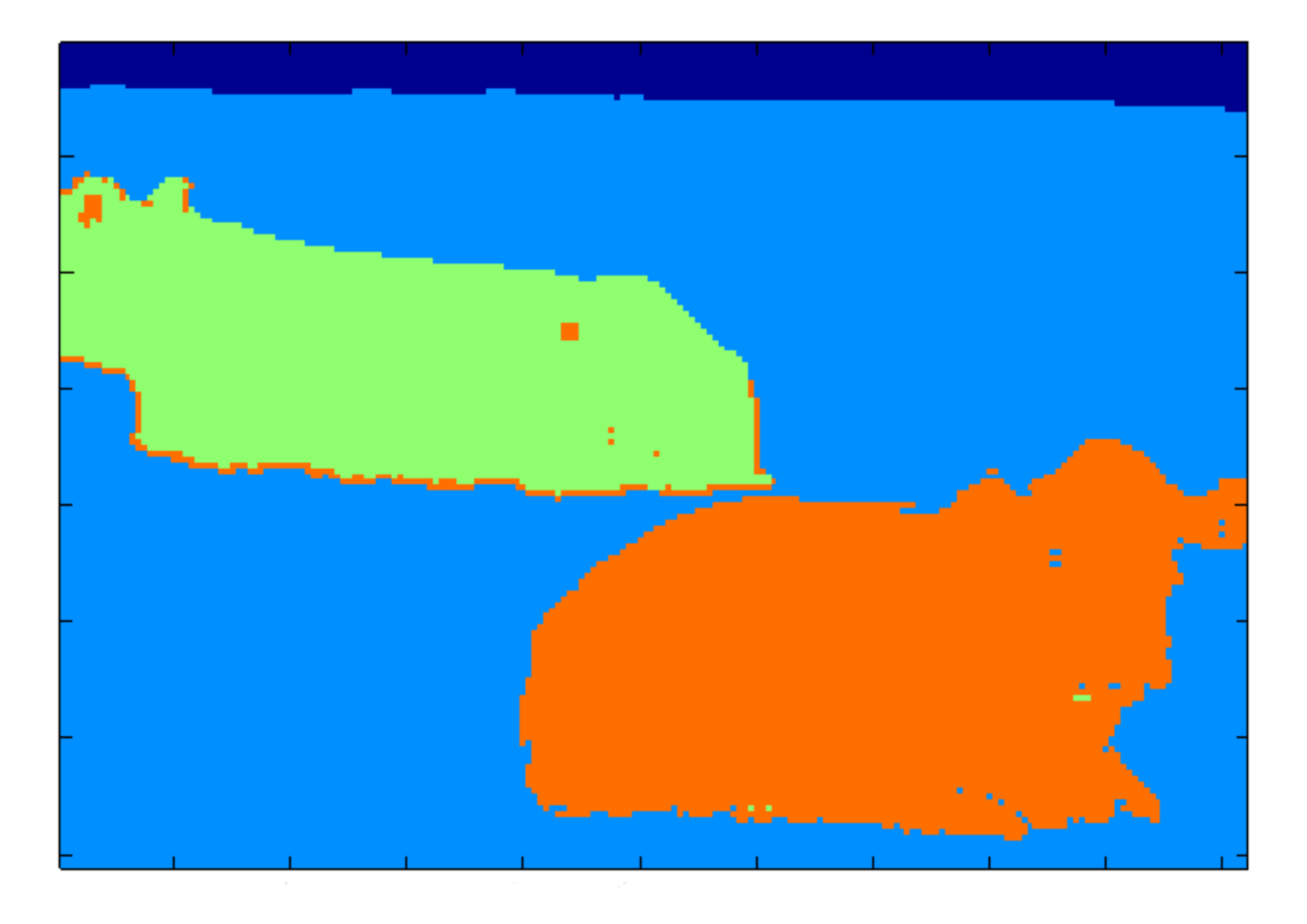}}\\
\subfigure[]{\includegraphics[width=0.231\textwidth]{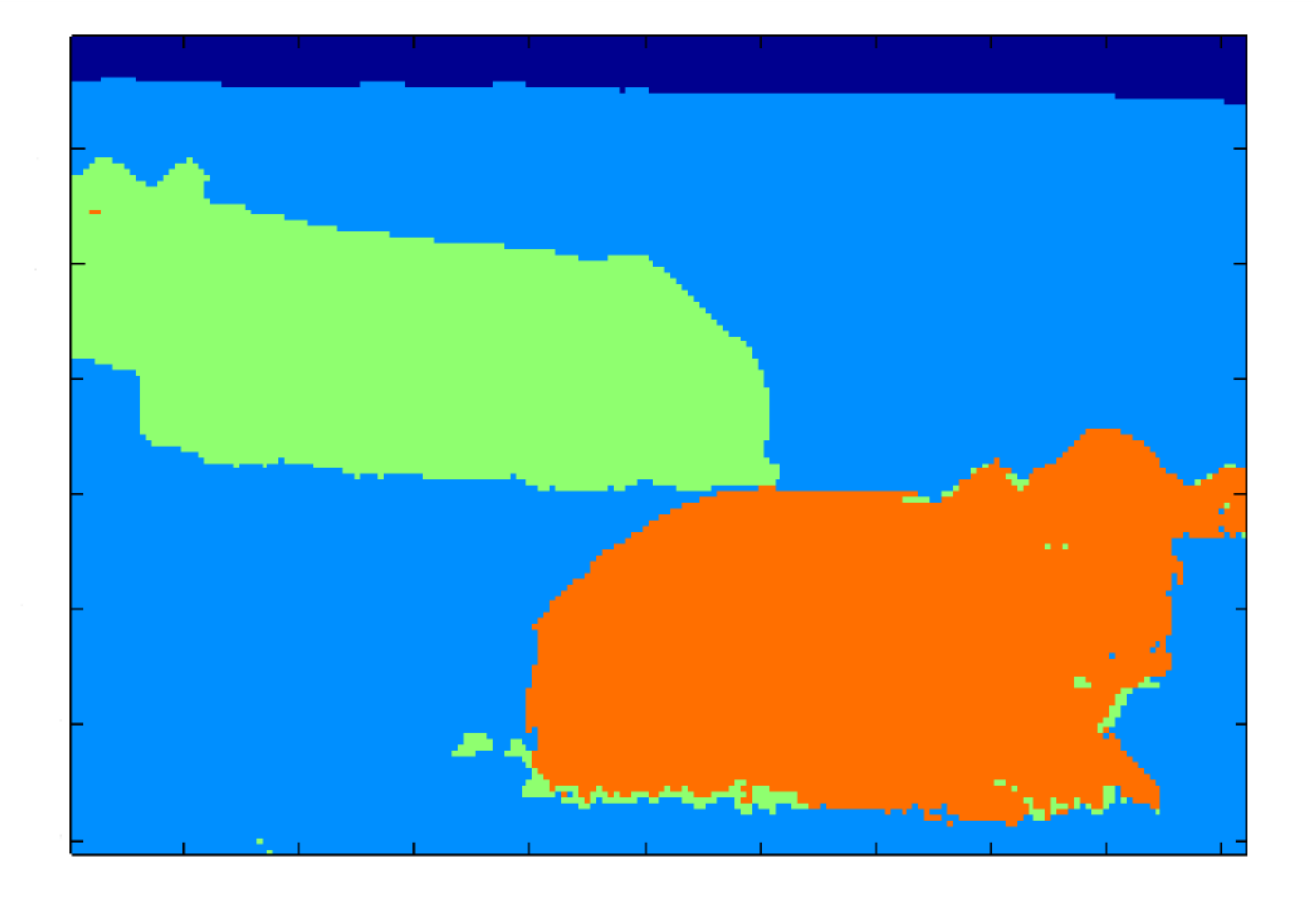}}
\subfigure[]{\includegraphics[width=0.231\textwidth]{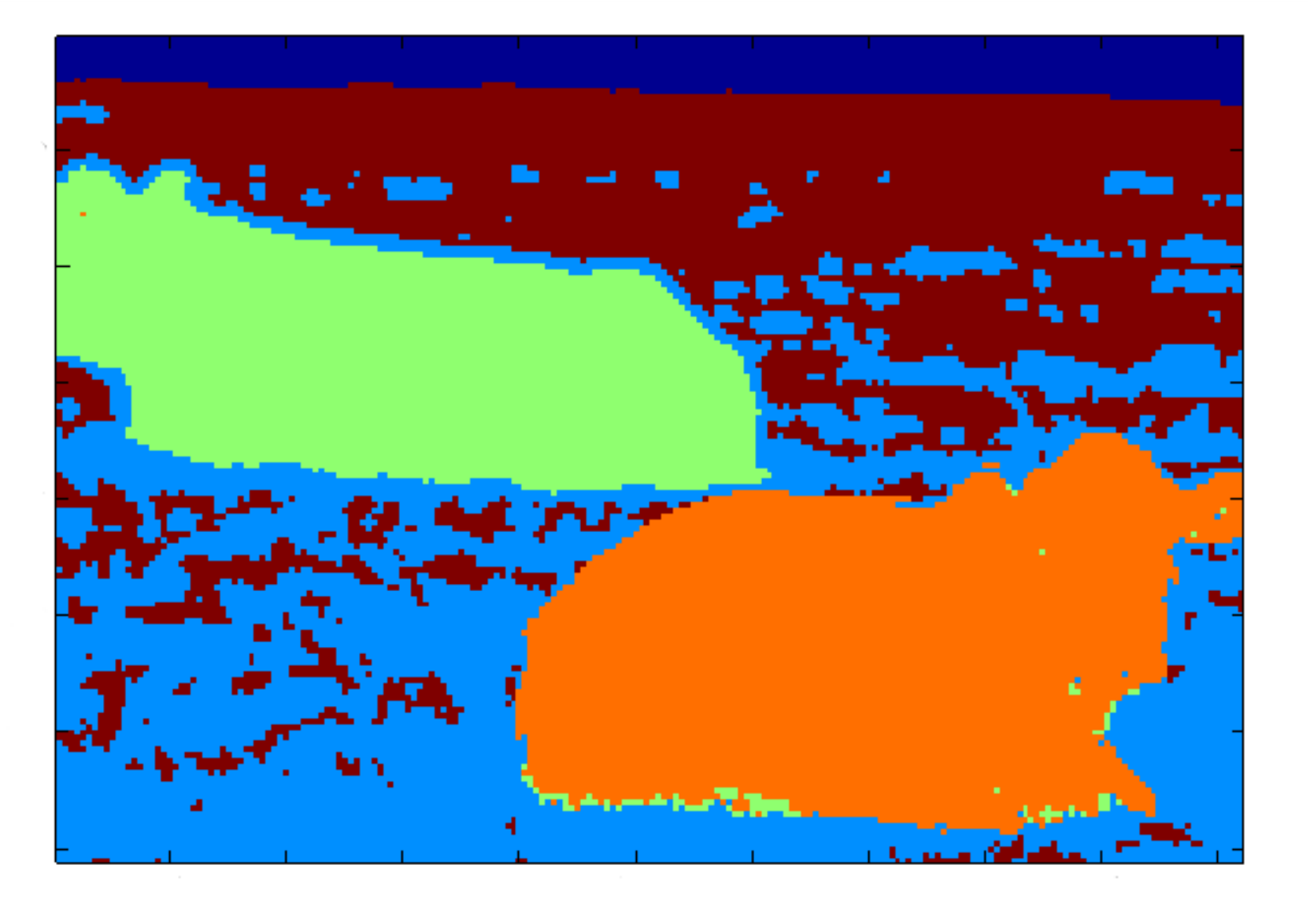}}
\caption{(a) Segmentation of the cow image in Fig.~\ref{schematic}. (b) obtained using optimization of multislice modularity with interslice coupling parameter $\omega = 0.3$.  The horizontal axis shows the slice index $s \in \{1,\ldots,6\}$, which has an associated resolution-parameter value of $\gamma_s = 0.04s - 0.03$.  The vertical axis gives the sorted pixel index.  Color in each vertical stripe indicates the community assignments of the pixels in the corresponding network slice. We also show the segmentation that we obtain in the images for (b) $\gamma_s = 0.05$, (c) $\gamma_s = 0.13$, and (d) $\gamma_s = 0.21$.
}
\label{cow}
\end{figure}
%======================================================================================================================
%Data and Results--LAPD
\section{Data and Results}
\subsection{LAPD Field Interview Data}
In \cite{ICDM}, we used data with both geographic and social information
about stops involving street gang members in the Los Angeles Police Department (LAPD) Division of Hollenbeck \cite{reu}.  We optimized multislice modularity (\ref{multimod}) as a means of unsupervised clustering of individual gang members without prior knowledge of the number of gangs or affiliation of the members.  We subsequently examined network diagnostics over slices to attempt to estimate the number of gangs that is stable across multiple resolution-parameter values and that also corresponds roughly to the number expected by the LAPD.
%======================================================================================================================
%Cow Images
 \subsection{Cow Image}
We segment the cow image in Fig.~\ref{schematic}(b) (which contains about $3\times 10^4$ pixels) without specifying the number of image components. We build a graph of this image in which each node corresponds to a pixel and each edge indicates the similarity between a pair of pixels.  We associate a $3\times 3$ pixel-neighbor patch with each pixel $i$ in the image. Let $p_D(i,j)$ denote the $L_2$ norm of the difference of patches corresponding to nodes $i$ and $j$. The adjacency matrix $\bf{A}$ that we use in each layer of the multislice network has elements
 \[
 	A_{ij} = \exp\left\{ \frac{-p_D^2(i,j)}{\tau(i) \tau(j)} \right\}\,,
\]
 where $\tau(i)$ is the 30th smallest $p_D$ between pixel $i$ and other pixels \cite{localscaling}.
We construct a multislice network that consists of six copies of ${\bf A}$.  We associate the resolution parameter value $\gamma_s = 0.04s - 0.03$ with slice $s \in \{1,\ldots,6\}$.  We then optimize multislice modularity and obtain the image segmentations shown in Fig.~\ref{cow}.  (Color indicates group assignments.)  With this procedure, we are able to identify all four components of the image.  As indicated in panel (a), we obtain smaller-scale communities (i.e., groups of pixels) as we increase the value of the resolution parameter.  Importantly (see the discussion in Section \ref{methods}), the coupling between slices enforces some consistency in clustering identical nodes similarly across slices. In panels (b) and (c), we observe a good segmentation of the two cows, the sky, and the background grass. As indicated in panel (d), the three groups corresponding to the two cows and the sky stay relatively stable, but the group corresponding to the grass breaks down by the sixth slice.\\

This application on image segmentation is computationally expensive due to the large number of pixels. It takes a lot of computational memory and time to run the optimization using more slices, which we would like to do in order to investigate how the segmentation evolves over a larger range of resolution values.  Computational improvements will be necessary to conduct more detailed analysis.

%{\bf map: Huiyi, are you using GenLouvain v 1.2?  that is much faster.  though I recall from our discussions that other steps besides genlouvain were the rate-limiters}

%======================================================================================================================

\section{Future Directions}

As mentioned above, optimization of multislice modularity can be computationally expensive. As the size of network data has increased tremendously, it is crucial to develop efficient algorithms to cluster network nodes to obtain insights on applications like social networks and images. To do this, one needs to take advantage of data sparsity to help speed up optimization processes.  %It is also important to characterize and analyze the performance of methods such as modularity optimization.
Aside from
the computational cost, how to characterize and analyze the
performance of modularity optimization is of importance as well.

%========================================================================================================================
% conference papers do not normally have an appendix

% use section* for acknowledgement
\section*{Acknowledgments}
We are grateful to the LAPD Division of Hollenbeck, and Megan Halvorson, Shannon Reid, Matt Valasik, James Wo, and George E. Tita, at the Department of Criminology, Law, and Society of UCI, for the collection, digitization, and cleaning, of the LAPD Field Interview data. We also thank P. Jeffrey Brantingham for educating us about the anthropology of gangs. This work is supported by 
ONR grant N000141210838, ONR grant N000141210040,
AFOSR MURI grant FA9550-10-1-0569, NSF grant DMS-0968309 and ONR grant
N000141010221. MAP acknowledges a research award (\#220020177) from the James S. McDonnell Foundation, and he thanks Andrea L. Bertozzi for hosting his visit to UCLA.
%======================================================================================================================
%Acknowledgements to be added to camera ready version upon acceptance.

% trigger a \newpage just before the given reference
% number - used to balance the columns on the last page
% adjust value as needed - may need to be readjusted if
% the document is modified later
%\IEEEtriggeratref{8}
% The "triggered" command can be changed if desired:
%\IEEEtriggercmd{\enlargethispage{-5in}}
%
% references section

% can use a bibliography generated by BibTeX as a .bbl file
% BibTeX documentation can be easily obtained at:
% http://www.ctan.org/tex-archive/biblio/bibtex/contrib/doc/
% The IEEEtran BibTeX style support page is at:
% http://www.michaelshell.org/tex/ieeetran/bibtex/
\bibliographystyle{IEEEtran}
% argument is your BibTeX string definitions and bibliography database(s)
\bibliography{IEEEabrv,bibHuiyi,icdm}
%
% <OR> manually copy in the resultant .bbl file
% set second argument of \begin to the number of references
% (used to reserve space for the reference number labels box)

%%%%%%%%%%%%%

% that's all folks
\end{document}